\documentclass[12pt, margin = 1in]{article}

\usepackage[inline,shortlabels]{enumitem}
\usepackage{float}
\usepackage{mathtools}
\usepackage{bm}
\usepackage{subdepth}
\usepackage{graphicx}
\usepackage{pdflscape}
\usepackage[round]{natbib}
\usepackage{tikz}
\usetikzlibrary{arrows.meta}
\usetikzlibrary{decorations.markings}
\usepackage[margin=1in]{geometry}
\usepackage[linesnumbered]{algorithm2e}
\RestyleAlgo{ruled}

\tikzset{+ /.tip = {Bar[sep=-3pt 2,width=3pt 4]_[sep=0]}}
\usepackage[english]{babel}
\usepackage{longtable}
\usepackage{color}
\usepackage{mathtools}
\usepackage{amssymb,amsmath,amsthm}
\usepackage{multirow}
\usepackage[titletoc,title]{appendix}
\usepackage{authblk}
\usepackage{bbm}
\usepackage{setspace}
\usepackage{dsfont}
\usepackage[OT1]{fontenc}
\usepackage{subcaption}
\usepackage{refcount}
\usepackage{booktabs}
\graphicspath{ {../simulations/01_init_manuscript/graphs/} }
\usepackage{enumitem}

\usepackage{xr}
\usepackage[colorlinks,citecolor=blue,urlcolor=blue]{hyperref}

\usepackage[inline]{trackchanges}

\newtheorem{remark}{Remark}
\newtheorem{proposition}{Proposition}

\AtEndDocument{\refstepcounter{theorem}\label{finalthm}}
\AtEndDocument{\refstepcounter{equation}\label{finaleq}}

{
  \theoremstyle{definition}
  
}
{
  \theoremstyle{definition}
  
}
{
  \theoremstyle{definition}
  
}

\AtEndDocument{\refstepcounter{lemma}\label{finallemma}}

\renewcommand{\P}{\mathsf{P}}

\newcommand{\E}{\mathsf{E}}
\renewcommand{\P}{\mathsf{P}}

\newcommand{\V}{\mathsf{Var}}

\usepackage{accents}

\DeclarePairedDelimiterX{\norm}[1]{\lVert}{\rVert}{#1}

\pgfdeclarelayer{background}
\pgfsetlayers{background,main}
\usetikzlibrary{arrows,positioning}
\tikzset{
>=stealth',
punkt/.style={
rectangle,
rounded corners,
draw=black, very thick,
text width=6.5em,
minimum height=2em,
text centered},
pil/.style={
->,
thick,
shorten <=2pt,
shorten >=2pt,}
}
\newcommand{\Vertex}[3]% pos, name
{\node[minimum width=0.6cm,inner sep=0.05cm] (#2) at (#1) {#3};
}
\newcommand{\Vertexr}[3]% pos, name
{\node[rectangle, draw, minimum width=0.6cm,inner sep=0.05cm] (#2) at (#1) {#2};
}
\newcommand{\ArrowR}[3]%
{ \begin{pgfonlayer}{background}
\draw[->,#3] (#1) to[bend right=30] (#2);
\end{pgfonlayer}
}
\newcommand{\ArrowLW}[3]%
{ \begin{pgfonlayer}{background}
\draw[->,#3] (#1) to[bend left=30] (#2);
\end{pgfonlayer}
}

\newcommand{\ArrowL}[3]%
{ \begin{pgfonlayer}{background}
    \draw[->,#3] (#1) to[bend left=45] (#2);
  \end{pgfonlayer}
}

\newcommand{\EdgeL}[3]%
{ \begin{pgfonlayer}{background}
\draw[dashed,#3] (#1) to[bend right=-45] (#2);
\end{pgfonlayer}
}

\newcommand{\Arrow}[3]%
{ \begin{pgfonlayer}{background}
\draw[->,#3] (#1) -- +(#2);
\end{pgfonlayer}
}

\allowdisplaybreaks
\pgfarrowsdeclare{arcs}{arcs}{...}
{
  \pgfsetdash{}{0pt} % do not dash
  \pgfsetroundjoin   % fix join
  \pgfsetroundcap    % fix cap
  \pgfpathmoveto{\pgfpoint{-5pt}{5pt}}
  \pgfpatharc{180}{270}{5pt}
  \pgfpatharc{90}{180}{5pt}
  \pgfusepathqstroke
}
\newcommand{\ArrowB}[3]%
{ \begin{pgfonlayer}{background}
    \draw[|-arcs,line width=0.4mm,shorten <= 0.3cm,shorten >= 0.3cm,#3] (#1) -- +(#2);
  \end{pgfonlayer}
}
\newcommand{\independent}{\perp \!\!\! \perp}

\newcommand{\titlepaper}{Transportability of aggregate trial results to an
external environment in causally interpretable
meta-analysis}

\date{\today}
\author[1,$\dagger$]{\small Tran Trong Khoi Le}
\author[1,$\dagger$]{\small Marie-Félicia Béclin}
\author[1]{\small Sivem Afach}
\author[1]{\small Tat Thang Vo \footnote{\small \textbf{Correspondence:} \\ Tat Thang Vo, Research Group EPIDERME, Faculty of Medicine, University Paris Est Creteil, France\\Email: tat-thang.vo@u-pec.fr}}
\affil[1]{\small EPIDERME, INSERM U955, Institute Mondor of Biomedical Research, University Paris Est Créteil, France}

\usepackage[colorinlistoftodos,textwidth=2.1cm]{todonotes}

\usepackage{xcolor}

\definecolor{lightblue}{rgb}{0.68,0.65,0.95}
\newcommand{\answer}[1]{\textcolor{black}{ #1}}

\title{\titlepaper}
\begin{document}
\maketitle
\begin{abstract}
    In evidence synthesis, multilevel modeling approaches (MMAs) are commonly employed to combine aggregate data (AD) and individual participant data (IPD). These approaches rely on an aggregate outcome model that is ideally obtained by integrating the prespecified individual-level outcome model over the covariate distribution observed in each eligible study. In non-linear settings, such an integration may however be analytically intractable and requires approximations. In this paper, we propose a novel method for incorporating AD into causal meta-analysis of IPD studies that can overcome this challenge. Rather than relying on an aggregate outcome model that is difficult to be correctly formulated, we propose modeling the trial membership as a function of baseline covariates. This model allows one to estimate the individual-level outcome model in each AD study by leveraging IPD available in other trials, and then to transport the treatment effects estimated from both AD and IPD trials to an external target population, even when only aggregate covariate data are available for that population. Unlike previous proposals, we do not require pseudo-IPD to be generated from the aggregate data, which helps minimize bias due to incomplete information on the covariate distribution in each AD trial and in the target population. 
\end{abstract}
\section{Introduction}
Meta-analysis methods involve combining and analyzing quantitative evidence from multiple studies to produce results based on a whole body of research. Traditional meta-analysis methods typically rely on aggregated data (AD) extracted from published reports to estimate the mean of the average treatment effects across studies. An alternative but increasingly popular approach is meta-analysis of individual participant data (IPD), in which the raw data of each trial are obtained and directly synthesized \citep{debray2015get, kawahara2018meta, smith2016individual}. 

\answer{Despite its methodological advantages, rigorous implementation of meta-analysis remains challenging due to two key obstacles: fragmented data availability and limited external validity. Although IPD meta-analysis enables more flexible modeling and generally yields higher-quality evidence, IPD are rarely available for all relevant studies \citep{rogozinska2017meta}. As an example, a recent survey in vascular and cognitive medicine found that nearly 47\% of contacted trial investigators did not respond to IPD requests, and 28\% explicitly declined to share data \citep{scutt2020data}. Similarly, in a large oncology evidence synthesis effort, IPD were unavailable for about 17\% of trials \citep{fayard2018impact}. Consequently, a substantial portion of the evidence base often resides in studies accessible only through AD. Methods to integrate IPD and AD are therefore essential to fully capture the available evidence and to mitigate the selection bias that may arise when analyses are restricted to a non-representative subset of trials. }

\answer{A second challenge concerns the gap between research settings and clinical practice. Randomized trials typically employ strict inclusion criteria that have patient samples different from the diverse profiles of real-world target populations. As a result, average treatment effects estimated within trials may not be directly relevant for clinical or policy decision-making. To address this limitation, recent work has introduced the framework of \textit{causally interpretable} (or \textit{population-adjusted}) meta-analysis \citep{vo2019novel,dahabreh2020toward,berenfeld2025causal}. This two-stage approach first standardizes trial-specific estimates to a target population and then synthesizes the standardized effects across studies, thereby improving clinical relevance. However, existing implementations of this framework typically require full access to IPD from all eligible trials, as well as from the target population, requirements that are often impractical in real-world settings.}

During the last few years, many statistical approaches have been proposed to combine AD and IPD when pooling trial results in standard and causal meta-analysis \citep{ravva2014linearization, upreti2019model, sutton2008meta, riley2010meta,pigott2012combining, riley2007evidence, vo2025integration, phillippo2020multilevel}. Among these approaches, those based on multilevel models are commonly used to examine the treatment effects in different patient subgroups. These approaches recognize the existence of natural hierarchy (population-level versus individual-level) in the available data, and propose separate model for each level \citep{sutton2008meta,upreti2019model}. Nonetheless, relationships developed between predictors and responses in AD models may not reflect individual predictor–response relationships, leading to ecological bias \citep{ravva2014linearization, upreti2019model}. To overcome this bias, the AD model should ideally be derived by marginalizing the underlying individual-level model over the covariate distribution of each study. However, such derivations are often complex and typically require approximations \citep{phillippo2020multilevel, phillippo2024multilevel}, since exact derivation of the AD model is only analytically feasible when all baseline covariates are discrete \citep{jackson2006improving, jansen2012network}, or when the individual-level outcome follows a simple linear model \citep{ravva2014linearization}. 
%For time-to-event outcomes that obeys non-linear models on the individual level, the AD model is untractable and the approximation of its likelihood by computational methods requires additional assumptions on the covariance matrix of covariates that is not fully observed in each AD trial \citep{phillippo2024multilevel}. 
In addition, methods based on multilevel modeling often rely on the assumption that the relative effects conditional on baseline covariates are constant across trials and in the target population \citep{phillippo2020multilevel,phillippo2024multilevel}. This assumption can be violated when there exists treatment version heterogeneity across studies, which is unavoidable in practice. For example, trials may differ in non-compliance rates or evaluate pharmacological interventions that vary in dose, route of administration, and other characteristics. To account for such heterogeneity, one approach is to assume that the main effects of the treatment in different individual-level outcome models are random draws from a hypothetical normal distribution, and to estimate the mean of this distribution. This strategy, however, does not allow one to obtain marginal treatment effect estimates for a clearly defined target population \citep{vo2019novel, dahabreh2020toward}. 

In this paper, we introduce a novel method for incorporating aggregate data into causally interpretable meta-analysis that can address the above limitations of multilevel modeling techniques. This method draws inspiration from propensity score-based methods in the causal inference literature \citep{vo2019novel, zubizarreta2023handbook}, as well as matching-adjusted indirect comparison (MAIC) methods in health technology assessment \citep{signorovitch2012matching, truong2023population}. Instead of employing an aggregate outcome model that is difficult to be correctly formulated, we alternatively model the trial membership as a function of baseline covariates. Importantly, the individual-level outcome model and the trial membership model can be specified in an orthogonal manner, i.e. the functional form of one does not constrain the other. 
The model for trial membership enables one to estimate the individual-level outcome model in each AD study, via leveraging IPD available in another eligible trial. The estimated outcome model of the AD trial is then  marginalized over the covariate distribution in the target population, either via G-computation if IPD is available for the target population,  or via a second stage of inverse weighting when only summary-level covariate data are available in that population.

\answer{Methodologically}, our proposed framework offers some potential advantages over existing approaches to combine IPD and AD. First, it avoids the need to generate pseudo-IPD from available AD--a step that can be particularly challenging and requires strong assumptions when information on the joint distribution of covariates is incomplete in the AD trials or in the target population \citep{phillippo2020multilevel,rott2024causally}. Second, by modeling trial membership rather than the aggregate outcome, we bypass the need to evaluate or approximate complex integrals that arise when deriving aggregate outcome models in non-linear settings. Third, the method can accommodate heterogeneity in treatment effects across studies and does not impose strict assumptions on effect constancy across trials. Fourth, it remains applicable even when the target population is external to all trial populations. \answer{Existing methods, in contrast, are
constrained by one or more of these challenges. For example, many approaches require the target population to coincide with that of an eligible trial \citep{vo2025integration}, while others rely on additional prior information, such as the outcome prevalence in specific subgroups of untreated individuals in the external population \citep{phillippo2020multilevel}. The proposed method does not depend on such information, thereby broadening its applicability in practice}.  

The paper is organized as follows. In the next section, we introduce the proposed weighting-based method and provide a rigorous analysis of its asymptotic properties. We then evaluate the finite-sample performance of the method using simulated data, and apply it to a real-world meta-analysis assessing the comparative effectiveness of various biologic treatments for psoriasis. The final section concludes and proposes directions for future research.
\section{Estimating conditional and marginal treatment effect under treatment version heterogeneity and limited access to data}
\subsection{Causal estimands and identification}
Assume that one wants to meta-analyze results of $K$ randomized studies $S=1,\ldots,K$ that evaluate the effect of a binary treatment $X$ on a binary outcome $Y$. Across all studies, individual-level data on a set of baseline characteristics $\bm L$ are also collected. The trials are deemed as sufficiently similar to be meta-analyzed, despite some heterogeneity in the distribution of covariates $\bm L$ and in the version of treatment and control evaluated across studies (\answer{e.g. route of administration, dosage or standard of care, etc}). In addition, the IPD on $(\bm L,X,Y)$ are accessible in studies $S = Z+1, \ldots, K$, but not in studies $S=1,\ldots, Z$, where $Z<K$. We thus observe $n$ trial observations $O_i$ that are independent and identically distributed, where $O_i=\big(L_iI(S_i\ge Z),X_iI(S_i\ge Z),Y_iI(S_i\ge Z),S_i\big)$, with $i=1,\dots,n$ and $n$ being the total number of patients across trials. Despite no direct access to IPD, the following summary statistics are available from trials $j=1,\ldots, Z$, for $x=0,1$:
\begin{enumerate}
\item[(D1)] \label{S.i} The sample mean and sample variance of each component $L_t$ of $\bm L$, i.e. $\hat \E (L_t\mid X=x,S=j)$ and $\hat{\V}(L_t\mid X=x,S=j)$.
\item[(D2)] The outcome (sample) mean in each treatment arm, i.e. $\hat \E(Y\mid X=x,S=j)$.
\item[(D3)] \label{iii} The chance of receiving treatment $X=x$, i.e. $\P(X=x|S=j)$, and the number of patients $n_{xj}$ in treatment arm $X=x$.
\end{enumerate}
Let $Y(x,s)$ denote the potential outcome that would be observed under treatment level $x$ as implemented in study $S=s$. 

%Our first aim is to estimate the conditional odds ratio:
%\[
%\theta_j(\bm L)=\frac{\mathrm{odd}\{Y(x=1,j)=1\mid \bm L\}}{\mathrm{odd}\{Y(x=0,j)=1\mid \bm L\}} \quad \quad \mathrm{for~}j=1,\ldots,K,
%\]
%which describes the odds of the event under treatment version $j$ versus control version $j$ given baseline characteristics $\bm L$. Other scale such as risk difference and risk ratio can also be considered. While subgroup treatment effects such as $\theta_j(\bm L)$ can help facilitate decision making on an individual level, marginal (or population-level) treatment effects can be of more interest for decision making in health technology assessment. 
Our aim is to estimate the marginal log odds ratio:
\[
\theta_s^0=\log \frac{\mathrm{odd}\{Y(x=1,s)=1\mid S=0\}}{\mathrm{odd}\{Y(x=0,s)=1\mid S=0\}} \quad \quad \mathrm{for~}s=1,\ldots,K.
\]
Here, with a slight abuse of notations, we denote $S=0$ as an external target population in which no trial has been previously conducted, i.e. only data on $\bm L$ are available for this population. Intuitively, $\theta_s^{0}$ represents the average treatment effect on the log odds ratio scale that would potentially be observed in population $S=0$ when comparing treatment version $s$ to control version $s$. As noted by \citet{phillippo2020multilevel}, the individual-level data on $\bm L$ in the target population may be inaccessible to the analyst. Instead, the following summary statistics are available from $S=0$:
\begin{itemize}
    \item  [(D4)] The sample mean and sample variance of each component $L_t$ of $\bm L$, i.e. $\hat \E (L_t\mid S=0)$ and $\hat{\V}(L_t\mid S=0)$, obtained from a representative sample $\mathcal{S}_0=\{i:S_i=0\}$ of size $n_0$ drawn from the target population. 
\end{itemize}

To identify $\theta_s^{0}$ from the observed data, we adopt the same set of causal assumptions often made in the transportability literature, namely:
\begin{itemize}
   \item [(A1)] \textit{Transportability}, i.e., $Y \left( x,s \right) \independent S\mid \bm L$ where $s=0,\ldots,K$, which requires that $\bm L$ contains all outcome prognostic factors that are differentially distributed \answer{between the target population and each trial population.}
    \item [(A2)] \textit{Positivity}, i.e. $\P\{0<\P \left( S=s \mid  \bm L \right) <1\} = 1$, where $s=0,\ldots,K$, which requires that there is an adequate overlap in the distribution of $\bm L$ \answer{between the target population and each trial population, and between two trial populations.}
    \item [(A3)] \textit{Consistency}, i.e. $Y \left( x,s \right) =Y$ when $S=s$ and $X=x$, which states that $Y (x,s)$ agrees with the observed outcome $Y$ for all individuals in study $s$ receiving treatment $x$, where $s=1,\ldots,K$. 
    \item [(A4)] \textit{Ignorability within each trial}, i.e. $Y \left( x, s \right) \independent X \mid S=s$ where $s=1,\dots, K$, which holds when the treatment is randomized within each study.
\end{itemize}
Under these assumptions, it has been proven that:
\begin{align*}
    \E\{Y(x,s)\mid S=0\} &= \sum_l\E(Y\mid X=x,\bm L=\bm l,S=s)\times \P(\bm L=\bm l\mid S=0)
\end{align*}
which allows one to identify $\theta_s^{0}$ \citep{vo2019novel,vo2025integration}.   
Based on this identification result, in the transportability and \answer{health technology assessment} (HTA) literature, multiple approaches have been developed to estimate $\theta_s^{0}$ when the IPD on $(\bm L, X,Y)$ are available in the source trial $S=s$, and either the IPD or the AD of $\bm L$ are available in the target population $S=0$ \citep{vo2019novel,dahabreh2020toward,signorovitch2012matching}. We contribute to this literature by developing estimation approaches for $\theta_s^{0}$ when the IPD are not available in the source trial $S=s$ and possibly not available in the target population $S=0$.

\subsection{Stage 1: estimating conditional treatment effects}
We first specify the following logistic outcome model for $s=1,\ldots,K$:
\begin{align} \label{mod1}
\E(Y|X, \bm L, S=s) = q(X,\bm L, \bm \phi_s)
&=\mathrm{expit}\{\phi_{0s} + \phi_{1s}X + \bm \phi_2^{\top}\bm L + \bm \phi_3^{\top}X\bm L )\} 
\end{align} 
Here, $\phi_{0s}$ and $ \phi_{1s}$
are the intercept and treatment coefficient in study $S=s$. These coefficients are heterogeneous across trials to reflect the variability in the versions of treatment and control evaluated in eligible studies. In contrast, $v(X, \bm L, \bm \phi_c) = \bm \phi_2^{\top}\bm L + \bm \phi_3^{\top}X\bm L$ reflects the part of the outcome model that is homogeneous across populations, where $\bm\phi_c = (\bm\phi_2,\bm\phi_3)$. 
Note that model (\ref{mod1}) can be easily estimated when $s=Z+1,\ldots,K$, due to the availability of the IPD in these trials. The main challenge thus lies in estimating $(\phi_{0s},\phi_{1s})$ for $s=1,\ldots,Z$. Before we address this challenge, some remarks are noteworthy here.

%\begin{remark}
  \answer{First, model (\ref{mod1}) is commonly adopted in so-called one-stage approaches to full IPD meta-analysis, i.e. when all trials have IPD accessible. These approaches typically posit a distribution for the trial-specific parrameters $(\phi_{0s},\phi_{1s})$, e.g. $(\phi_{0s},\phi_{1s}) \sim \mathcal{N}(\bm \Phi,\bm\Sigma)$, and aim to estimate the parameters $\bm \Phi$ and $\bm \Sigma$ of this distribution (e.g. by using restricted maximum likelihood). The method we propose below enables the estimation of such a random-effect model when the IPD are not available for all trials. }
%\end{remark}

%\begin{remark}
 \answer{
    Second, model (\ref{mod1}) can be further relaxed to allow for between-trial variability in additional coefficients (such as in the interaction effect $\phi_3$). Estimation of this more general model requires a richer set of summary statistics to be available in the AD trials $S=1,\dots,Z$. We discuss this extension at the end of this section.}
%\end{remark}

%\begin{remark}
 \answer{
    Finally, it is worth noting that to overcome the challenge due to AD, classical meta-regression methods often rely on hierarchical modeling. For instance, with one-dimension covariate $L$, \citet{sutton2008meta} suggest the following hierarchical model:
\begin{align*}
    \E(Y\mid X, L,S=k) &= \mathrm{expit}\{\phi_{0k} + \phi_{1k}X + \phi_{2k}L + \phi_3 XL\}\\
    \E(Y\mid X, S=j) &= \mathrm{expit}\{\phi_{0j} + \phi_{1j}X + \phi_3X\E(L|S=j)\}\\
    \phi_{1s} &\sim \mathcal{N}(\phi_1, \tau^2)
\end{align*}
where $j=1,\ldots, Z;k = Z+1,\ldots, K$ and $s=1,\ldots, K$. However, relationships between $(X,L)$ and $Y$ inferred from AD in trial $j=1,\ldots, Z$ may fail to
reflect individual-level associations. In particular, it is generally infeasible to posit an individual-level outcome model $\E(Y|X,L,S=j)$ for an AD trial $j$ such that, after marginalizing over $L$, the resulting marginal expectation $ \E(Y|X,S=j)$ satisfies the above specification. To avoid so-called ecological bias,  \citet{phillippo2020multilevel} alternatively consider the following individual-level outcome model:
\begin{align} \label{ml-nmr}
    \E(Y\mid X, L,S=s) &= \mathrm{expit}\{\phi_{0
    s} + \phi_{1}X + \bm\phi_{2}^\top \bm L + \phi_3^\top X\bm L\}
    \end{align}
for $s=1,\ldots,K$, then approximate the Poisson binomial distribution of $E(Y\mid X,S=j)$ for $j=1,\dots,Z$ by a binomial distribution. Simultaneous ftting of IPD and AD under this framework to estimate $(\phi_1,\bm\phi_2,\bm \phi_3)$, however, requires the generation of pseudo-IPD for covariates $\bm L$ in AD trials. This is often impractical, as detailed information on the distribution of $\bm L$ is rarely available in AD trials. Moreover, this approach assumes no treatment effect heterogeneity across trials--specifically, that the conditional treatment effect satisfies \[\phi_s(\bm L) = \frac{\mathrm{odd}{(Y=1 \mid X=1, \bm L, S=s)}}{\mathrm{odd}{(Y=1 \mid X=0, \bm L, S=s)}} = \phi(\bm L) = \phi_1 + \bm\phi_3^\top \bm L.\] Consequently, it relies on an individual-level outcome model that is more restrictive than the model specified in (\ref{mod1}), which we consider in this work.
%Of note, in the exceptional setting of linear models, such an issue can be dealt with by deriving the aggregated data model based on the individual-level model, then simultaneously fitting both the AD and IPD. This approach, however, is not extensible to non-linear settings, as the resulting AD model will be overly complicated even for a very simple individual-level (non-linear) outcome model.
%\end{remark}
}

While the specification of multiple parametric models appears necessary to take into account the aggregate data from studies $S=1,\dots,Z$, prior approaches have primarily focused on modeling the relationship between aggregate outcomes and aggregate covariates. To date, little attention has been given to the possibility of modeling alternative aspects of the data-generating process. In what follows, we introduce a weighting-based strategy to estimate $(\phi_{0s},\phi_{1s})$ where $s=1,\ldots,Z$ that can overcome the above-mentioned limitations of hierarchical modeling approaches. Such a strategy is developed based on the following identity:
    \begin{align}\label{mod:mm2}\E\bigg\{ I(x,k)~\frac{\E(Y|x, \bm L, S=j)}{\P(X=x|S=k)}~\frac{\P(S=j|\bm L)}{\P(S=k|\bm L)}\bigg\} = \E(Y|x,S=j)~ \P(S=j) \end{align}
Here, $I(x,k)$ is the short-hand notation for $I(X=x,S=k)$, where $k\in \{Z+1,\ldots,K\}$ is an IPD trial and $j\in \{1,\ldots,Z\}$ is an AD trial. A formal proof is provided in the Online Supplementary Materials.
\answer{Intuitively, this identity formalizes the idea that appropriate reweighting by treatment assignment and study membership makes it possible to construct a synthetic version of study $j$ using individuals from study $k$ (as represented by the left-hand side), thereby recovering the marginal outcome mean under treatment $x$ in study $j$ (as given in the right-hand side).}

The proposed identity further suggests that the coefficients $(\phi_{0j},\phi_{1j})$ specific to trial $j$ can be estimated by leveraging the IPD available in trial $k$. This is feasible provided that the common parameter $\bm\phi_c$ across populations and the chance of being in trial $j$ versus trial $k$ (given $\bm L$) can be consistently estimated. Such an approach is particularly appealing because it avoids the need to specify an outcome model for the AD as in hierarchical modeling approaches (thereby mitigating ecological bias), while still allowing AD to be incorporated in the estimation of $(\phi_{0j},\phi_{1j})$. The fitting procedure is described step by step below:

\textbf{Step 1 --} Estimate the parameter $\bm\phi_c = (\bm\phi_2,\bm\phi_3)^\top$ in model (\ref{mod1}) by using data from studies $k=Z+1,\ldots K$ for which IPD are available. 

One option (strategy A) is to fit a (correctly specified) logistic model within each IPD trial $k$, using standard maximum likelihood. This yields $K-Z$ estimates $\hat{\bm\phi}^{(k)}_c$ for $\bm\phi_c$.  

Alternatively, a single pooled estimate of $\bm \phi_c$ may be obtained (strategy B) by solving the esstimating equations \[\sum_{i=1}^n \bm g(\bm O_i,\bm\lambda) = \bm 0,\] 
where $ \bm \lambda$ denotes the vector of all parameters indexing the outcome models for trial $k=Z+1,\ldots,K$ and $\bm g (\bm O, \bm \lambda)$ collects the corresponding score functions. Specifically, the score function associated with $(\phi_{0k},\phi_{1k})$ is:
        \[
        \bm g_k =  I(S=k)~\begin{pmatrix}
            1 & X
        \end{pmatrix}^\top~\big[Y - \mathrm{expit}\{\phi_{0k} + \phi_{1k}X + v(X, \bm L, \bm \phi_c)\big]\]
and the (aggregated) score function associated with $\bm\phi_c$ is:
\[\bm g_c = \sum_{k=Z+1}^K I(S=k)~\begin{pmatrix}
    \bm L^\top & X\bm L^\top
\end{pmatrix}^\top~\big[Y - \mathrm{expit}\{\phi_{0k} + \phi_{1k}X + v(X, \bm L, \bm \phi_c)\big]\]
Solving these equations yields a unified estimate of the parameter $\bm \phi_c$ common across IPD trials.

\textbf{Step 2 --} Assume the following propensity score (or weight) model:
    \begin{align}\label{mod:we}
    \frac{\P(S=j|\bm L)}{\P(S=k|\bm L)} = m(\bm L, \bm \beta_{jk}) = \exp(\beta_{jk}^{(0)} + \beta_{jk}^{(1)}\bm L)
    \end{align}
 where $j=1,\ldots, Z$ and $k=Z+1,\ldots, K$. \answer{Note that this model depends on both population $j$ and $k$.} The parameters $\bm \beta_{jk}$ can be estimated by solving the sample analogue of the following moment equation:
    \begin{align} \label{mod:mm}
    \E\big\{I(S=k)~\bm p(\bm L)~ m(\bm L, \bm\beta_{jk}) \big\} = \P(S=j) ~\E\big \{\bm p(\bm L)|S=j\big \}
    \end{align}
    where the vector function $\bm p(\bm L)$ is chosen such that a sample estimate of $\E\{\bm p(\bm L)|S=j\}$ is available in the aggregated data of trial $S=j$, e.g., $\bm p(\bm L) = \begin{pmatrix}
        1 & \bm L^\top
    \end{pmatrix}^\top $. In this way, only the IPD of trial $k=Z+1,\ldots, K$ is needed to construct the sample analogue of equation (\ref{mod:mm}). 
    
    This flexible moment-based method also allows for more complex parametric weight models. For instance, if $m(\bm L_i, {\bm\beta}_{jk}) = \exp(\beta_{jk}^{(0)} + \beta_{jk}^{(1)}L_1 + \beta_{jk}^{(2)}L_2 + \beta_{jk}^{(3)}L_1L_2)$, one may choose $\bm p(\bm L) = \begin{pmatrix} 1 & L_1 & L_2 & L_1^2 \end{pmatrix}^\top$. As long as the number of estimating equations is equal to or exceeds the number of parameters in $\bm\beta_{jk}$ (and the model is correctly specified), the proposed method can provide consistent estimates of $\bm\beta_{jk}$.
    
    For simplicity, we restrict the discussion below to the case where the covariates $\bm L$ enter the model (\ref{mod:we}) linearly. Nonetheless, all subsequent steps (including asymptotic inference) can be extended straightforwardly to more complex parametric weight models.

\textbf{Step 3 --} We are now ready to apply the moment equation (\ref{mod:mm2}) to estimate $\phi_{0j}$ and $\phi_{1j}$. 

Under strategy A at step 1, these parameters are obtained by solving the sample equations:
    \begin{align} \label{eq5}
        \frac{1}{n_j}\sum_{i:X_i=x,S_i=k} ~\frac{\mathrm{expit}\big\{\phi_{0j} + \phi_{1j}x + v(x, \bm L_i, \hat{\bm \phi}_c^{(k)})\big\}}{P_{xk}}~ m(\bm L, \hat{\bm{\beta}}_{jk}) = \hat \E(Y|x,S_i=j)
    \end{align} 
    for $j=1,\ldots, Z$ and $k = Z+1, \ldots, K$. For a given AD trial $j$, using the data from IPD trial $k$ yields an estimate $\hat{\phi}_{xj}^{(k)}$ for $\phi_{xj}$.
    %The estimates $\hat{\phi}_{xj}^{(k)}$ of $\phi_{xj}$ obtained from trial $k=Z+1,\ldots, K$ can then be averaged (and possibly weighted by the sample size of trials $k$) to obtain a final estimate $\hat{ \phi}_{xj}$ for $\phi_{xj}$, where $x=0,1$. 
    
    Under strategy B at step 1, one can instead plug in the pooled estimate $\hat{\bm\phi}_c$ in place of $\hat{\bm\phi}_c^{(k)}$ in equation (\ref{eq5}), producing multiple estimates $\hat{\phi}_{xj}^{(k)}$ for $\phi_{xj}$ in a similar fashion.%, then averaging these estimates as in strategy A. %Alternatively, the (simplified) GMM can be used to obtain directly a unique estimate for $\phi_{0j}$ and $\phi_{1j}$. More precisely, denote:
    %\[
    %h_{jkx} = h(\bm O, x, \hat{\bm \beta}_{jk}, \bm\phi_j^*, \hat{\bm\phi}_c) = I(x,k) ~q(x,\bm L,\bm \phi_j^*, \hat{\bm \phi}_c) ~m(\bm L, \hat{\bm \beta}_{jk}) - I(j,x)Y
    %\]
    %Then, with a slight abuse of notation, a unique estimate $\hat{\bm\phi}_j^*$ for $\bm\phi_j^*$ can be expressed as: \[\hat{\bm\phi}_j^* = \mathrm{argmin}_{\bm\phi_j^*}||n^{-1}\sum_i \bm h_j(\bm O_i, \bm\phi_j^*)||_2^2\] 
    %where $\bm h_j = \begin{pmatrix}
    %    h_{j(Z+1)x}^\top & h_{j(Z+1)(1-x)}^\top & \ldots & 
    %    h_{jKx}^\top & h_{jK(1-x)}^\top
    %\end{pmatrix}^\top$. Indeed, this optimization exercise only requires the IPD from trial $k=Z+1,\ldots, K$ and the sample mean $\hat{\E}(Y|x,S=j)$ from study $j=1,\ldots, Z$.
    %\end{enumerate}

The above procedure will provide consistent estimates for $(\phi_{0j},\phi_{1j})$ in the AD trials $j=1,\ldots, Z$, provided that the propensity score model (\ref{mod:we}) is correctly specified or at least approximates well the (unknown) study assignment in the meta-analysis. Importantly, as in other weighting-based approaches, the weights derived from model (\ref{mod:we}) can be used as a mean to evaluate the similarity in patient characteristics across studies, which is important to prevent the combination of trials that are too different in case-mix \citep{vo2019novel}. \answer{In the presence of extreme weights, there exists individuals with a particularly high chance to be recruited in some specific trials and not in the others, so-called limited covariate overlap between populations. In practice, extreme weights are primarily associated with a loss of precision. However, they can also induce bias. When overlap is poor, estimation relies disproportionately on a small number of individuals with very large weights, so that the estimating equations are effectively dominated by a few observations. In such settings, non-negligible finite-sample bias can arise even when the underlying weight model is correctly specified, as the empirical estimating equations no longer approximate well their population counterparts. Moreover, extreme weights amplify any residual model misspecification or numerical instability, further contributing to biased estimation of $(\phi_{0j}, \phi_{1j})$. This phenomenon has been documented in the causal inference literature (e.g., \citealp{vo2019novel}).} Importantly, insufficient overlap in the target population of eligible trials may easily go unnoticed in standard outcome regression-based meta-analysis methods, which often rely on substantial extrapolations when overlap is limited \citep{vo2019novel}.

\begin{remark}
    When other summary statistics are available in AD trials, the proposed approach can be extended to estimate a more flexible outcome model than (\ref{mod1}). For instance, for $x,y\in \{0,1\}$, suppose that the sample mean $\hat\E(\bm L_1\mid X=x,Y=y,S=j)$ of some categorical covariates $\bm L_1$ in $\bm L$ is reported in trials $j=1,\ldots,Z$, as a result of subgroup analyses assessing potential effect modification by $\bm L_1$.
    
    These additional data allow us to postulate and fit a more complex outcome model, which allows the effect of some covariates $\bm L_2$ in $\bm L$ to vary across studies:
    \begin{align*}
    \E(Y|X, \bm L, S=s)
    &=\mathrm{expit}\{\phi_{0s} + \phi_{1s}X + \bm \phi_{2s}^\top\bm L_2 + \bm \phi_{3s}^\top X\bm L_2 + \bm \phi_4^\top\bm L_3 + \bm \phi_5^\top X\bm L_3\}
    \end{align*} 
    Here, $\bm L_2$ may be the same as $\bm L_1$ or different. If they differ, the dimension of $\bm L_2$ should be no larger than that of $\bm L_1$. An identity analogous to (\ref{mod:mm2}) then holds:
\begin{align*}\label{mod:mm4}
&\E\bigg\{ I(x,k)~\bm L_1\frac{\P(Y=y\mid x,\bm L, S=j)}{\P(X=x\mid S=k)}~\frac{\P(S=j\mid \bm L)}{\P(S=k\mid \bm L)}\bigg\}\\
&= \E(\bm L_1\mid x,y,S=j)~\P(Y=y\mid x,S=j)~\P(S=j) \end{align*}
for $x,y\in \{0,1\}$; $j=1,\ldots,Z$ and $k=Z+1,\ldots K$. Based on this equality, a strategy analogous to that described above can be developed to estimate $(\phi_{0j},\ldots,\bm \phi_{3j})$. For brevity, details are omitted.
\end{remark}
\subsection{Stage 2: transporting treatment effects to the target population}
Once the outcome model $\E(Y\mid X,\bm L,S=s)$ has been estimated for all trials $s=1,\ldots,K$, a G-computation estimator $\hat \P_{xs}^{0}$ of $\P_{xs}^{0} = \E\{Y(x,s)\mid S=0\}$ can be readily obtained when the IPD on $\bm L$ is available from the target population. Specifically, such an estimator can be written as:
\[\hat \P_{xs}^{0} = 
\frac{1}{n_{0}}\sum_{i\in \mathcal{S}_0}\mathrm{expit}(\hat\phi_{0s} + \hat\phi_{1s}x + \hat{\bm \phi}_2^\top \bm L_i + \hat{\bm \phi}_3^\top x\bm L_i) 
\]
which in turn allows estimation of the marginal treatment effect $\theta_s^{0}$. % is then estimated as: \[\hat\theta_s^{0} =\log \{\hat \P_{x=1,s}^{0}/(1-\hat \P_{x=1,s}^{0})\} - \log \{\hat \P_{x=0,s}^{0}/(1-\hat \P_{x=0,s}^{0})\}. \]

Our primary interest, however, lies in the more challenging setting in which the individual-level data $\mathcal{S}_0$ is not available, but only the sample means and variances of 
$\bm L$ within $\mathcal{S}_0$ (as described in D4) are reported. In this case, the moment-based approach introduced in the previous section (see equation (\ref{mod:mm})) can be employed to estimate the propensity score model:
\begin{align} \label{mod:tp}
    \frac{\P(S=0\mid\bm L)}{\P(S=k\mid \bm L)} = m(\bm L, \bm \beta_{0k}) = \exp(\beta_{0k}^{(0)} + \beta_{0k}^{(1),\top}\bm L),
\end{align}
using the IPD of trial $k=Z+1,\ldots,K$. 

An estimator of $\P_{xs}^0$ can then be constructed by leveraging the IPD from trial $k$, noting that:
\begin{align} \label{mod:tpe}
\P_{xs}^0 = \frac{1}{\P(S=0)}\E\bigg\{I(S=k)~\E(Y\mid x,\bm L,S=s)~\frac{\P(S=0\mid \bm L)}{\P(S=k\mid \bm L)} \bigg\}
\end{align}
\answer{Intuitively, equation (\ref{mod:tpe}) shows that the counterfactual mean outcome under version $s$ of treatment $x$ in the target population can be recovered by reweighting individuals from an IPD trial $k$ so that their covariate distribution matches that of the target population $S=0$. The weight $\P(S=0 \mid \bm L)/\P(S=k \mid \bm L)$ shifts the trial-$k$ sample toward the target population, while $\E(Y \mid x,L,S=s)$ supplies the predicted outcome under version $s$ of treatment $x$. Averaging this reweighted quantity and normalizing by $\P(S=0)$ thus yields the marginal mean outcome that would be observed if trial 
$s$ were transported to the target population, even when the IPD of both trial $s$ and the target population are not available.}

Specifically, to transport results of an 
IPD trial $k=Z+1,\ldots,K$
to the target population, we compute:
\[\hat \P_{xk}^{0,k} = \frac{1}{n_0}\sum_{i:S_i=k}\mathrm{expit}(\hat\phi_{0k}^{(k)} + \hat\phi_{1k}^{(k)}x + \hat{\bm \phi}_2^{(k),\top} \bm L_i + \hat{\bm \phi}_3^{(k),\top} x\bm L_i)\times m(\bm L_i,\hat{\bm\beta}_{0k}) \]
And to transport results of an 
AD trial $j=1,\ldots,Z$ to the target population, we compute:
\[\hat \P_{xj}^{0,k} = \frac{1}{n_0}\sum_{i:S_i=k}\mathrm{expit}(\hat\phi_{0j}^{(k)} + \hat\phi_{1j}^{(k)}x + \hat{\bm \phi}_2^{(k),\top} \bm L_i + \hat{\bm \phi}_3^{(k),\top} x\bm L_i)\times m(\bm L_i,\hat{\bm\beta}_{0k}) \]
In both expressions, the trial-specific estimates $(\hat{\bm \phi}_2^{(k)},\hat{\bm \phi}_3^{(k)})$ may be replaced by the pooled estimator $(\hat{\bm \phi}_2,\hat{\bm \phi}_3)$ of $(\bm\phi_2,\bm\phi_3)$ when Strategy B is used at stage 1 of the analysis. 

Once these probabilities have been estimated, the marginal treatment effect $\hat\theta_s^0$ can be readily computed for each trial $s=1,\ldots,K$. Note that for the AD trials $j=1,\ldots, Z$, different estimates $\hat\theta_j^{0,k}$ of $\theta_j^0$ can be obtained by leveraging IPD from different trials $k=Z+1,\ldots, K$. These trial-specific estimates may then be combined, potentially using weights proportional to the sample sizes of the IPD trials, to obtain a final estimate $\hat\theta_j^0$ of $\theta_j^0$. 

\begin{remark}\label{remark2}
    In the absence of access to \answer{IPD from} $\mathcal{S}_0$, estimating $\theta_s^0$ via the weighting approach based on (\ref{mod:tpe}) offers notable advantages over existing methods in the literature. For instance, \citet{phillippo2020multilevel} and  \citet{rott2024causally} propose generating pseudo-IPD for $\bm L$ in $\mathcal{S}_0$, followed by G-computation on these reconstructed data to estimate $\theta_j^0$. As mentioned earlier, however, reconstruction of IPD is challenging due to the lack of information on cov$(L_a,L_b\mid S=0)$ for any pairs of covariates $L_a,L_b$ in $\bm L$. In addition, \citet{phillippo2020multilevel} require an estimate of $\E(Y\mid X=0,\bm L=\bm 0,S=0)$ (under the assumption of no treatment version heterogeneity), which may be unavailable if no prior studies has been conducted in the subgroup $\bm L=\bm 0$ of the target population. Such an estimate is not required by our proposed approach.
    
    Our proposal is also closely related to the recent work by \citet{vo2025integration}.  While their approach avoids explicitly parameterizing treatment version heterogeneity across studies in the outcome model, it is applicable only when the target population $S=0$ coincides with one of the trial populations. In contrast, the method proposed here allows trial results to be transported to an external target population, regardless of whether IPD are available in the trial and the target population, a setting that is not accommodated by \citet{vo2025integration}. 
\end{remark}
\subsection{Stage 3: Meta-analyzing transported treatment effects}
\answer{We now discuss how to combine the standardized effects $(\hat\theta_1^0, \ldots, \hat\theta_K^0)$ to obtain a summary treatment effect for the target population $S=0$. To motivate the proposed pooling strategy, we first briefly review standard meta-analysis models, which serve as a foundation for the new model used to summarize $(\hat\theta_1^0, \ldots, \hat\theta_K^0)$.} 

\answer{Let $\gamma_j=\log \{ \mathrm{odd}\{Y(1,j)=1|S=j\}/\mathrm{odd}\{Y(0,j)=1|S=j\}$ denote the (unstandardized) treatment effect in trial population $j$, for $j=1,\ldots, K$. In a conventional random-effects meta-analysis model, the estimator $\hat{\gamma}_j$ of $\gamma_j$ satisfies:
\begin{equation} \label{mod:re.conv}
\begin{aligned}
\hat{\gamma}_j &= \gamma_j + \epsilon_j, \\
\gamma_j &= \Gamma + \alpha_j,
\end{aligned}
\end{equation}
where $\epsilon_j \sim \mathcal{N}(0, \sigma_j^2)$ denotes sampling error
and $\alpha_j \sim \mathcal{N}(0, \tau^2)$ is a study-specific random effect. The collection $\{\epsilon_1, \ldots, \epsilon_K, \alpha_1, \ldots, \alpha_K\}$ is assumed to be mutually independent. In this model, the summary treatment effect $\Gamma$ represents the mean of the study-specific effects $\gamma_j$ and is typically estimated as a weighted average of the observed effects $\hat{\gamma}_j$. The variance component $\tau^2$, often referred to as \textit{heterogeneity variance}, captures between-study variability in $\gamma_j$ arising from methodological differences and heterogeneity in case mix across studies.} 

\answer{When effect modifiers are unevenly distributed across trials, the clinical relevance of $\Gamma$ becomes unclear, as the model leaves implicit the population for which the summary effect is intended to apply. In practice, clinical decision-making requires inference on treatment effects for well-defined target populations of patients, rather than averages taken over heterogeneous trial populations. In view of this, the use of population-standardized effect estimates, as pursued in this work, has become increasingly common in the literature on \textit{causally interpretable meta-analysis} \citep{vo2019novel,vo2025integration,dahabreh2020toward}.}

Once the standardized effect estimates $\hat{\bm\theta} =(\hat\theta_1^0, \ldots, \hat\theta_K^0)$ are obtained, one can derive the summary treatment effect in the target population $S=0$ by using a random-effect model analogous to (\ref{mod:re.conv}), i.e.:

\begin{equation} \label{mod:re}
\begin{aligned}
   \answer{\hat{\bm\theta}}&=  \answer{\bm\theta + \bm\eta }\\
    \bm \theta &=\Theta\cdot\mathbf 1_K+\bm \beta
\end{aligned}   
\end{equation}
where $\bm\beta \sim \mathcal{\bm N}(0,\zeta^2\mathbf I_K)$ is a random-effects vector; $\bm\eta\sim \mathcal{\bm N}(0,\bm\Sigma)$ is the random error vector; and $\bm\beta$ is independent of $\bm\eta$. Here, $\mathbf 1_K$ is a $K$-dimention vector of ones and $\mathbf I_K$ is the $K\times K$ identity matrix. The fixed-effect parameter $\Theta$ in this model represents the summary treatment effect in the target population $S=0$, while the heterogeneity variance $\zeta^2$ quantifies \textit{beyond case-mix heterogeneity}, i.e. the variability in the treatment versions $s=1,\ldots,K$ \citep{vo2019novel}. 

\answer{The proposed meta-analysis model can be fitted by using restricted maximum likelihood or Markov Chains Monte Carlo simulations. Despite its similarity to the
conventional random-effects model (\ref{mod:re.conv}), it is important to note that model (\ref{mod:re}) is not identical. The key distinction is that the standardized effect estimates $(\hat\theta_1^0, \ldots, \hat\theta_K^0)$ are generally correlated, unlike $(\hat\gamma_1,\ldots,\hat\gamma_K)$. As a result, the covariance matrix $\bm\Sigma$ for the stadardized effects is not diagonal, unlike that of the unstandardized effects. Consequently, model (\ref{mod:re}) cannot be fitted directly using off-the-shelf software designed for conventional random-effects meta-analysis.}

\subsection{Asymptotic inference}
We now investigate whether the asymptotic variance of $\hat{\bm\theta}$ can be estimated from the available data. 

Let $\bm \psi$ denote the vector of parameters of interest, which includes all parameters indexing the logistic outcome models and the weight models, as well as the probabilities $\P_{xs}^0$, for $x=0,1$ and $s=1,\ldots,K$. Let $\hat{\bm\psi}$ denote the estimator of $\bm\psi$ obtained from the proposed procedure. 

By M-estimation theory, 
\begin{align}\label{sw}
    \sqrt{n}(\hat{\bm\psi} - \bm\psi) \xrightarrow{d} \bm{\mathcal{N}}\bigg(\bm 0, \E\bigg[-\frac{\partial \bm f}{\partial\bm\psi}(\bm O, \bm\psi_0) \bigg]^{-1} \V\big[\bm f(\bm O, \bm \psi_0)\big] \E\bigg[-\frac{\partial \bm f}{\partial\bm\psi}(\bm O, \bm\psi_0) \bigg]^{-1,\top} \bigg)
\end{align}
where $\bm O = (\bm L,S,X,Y)$, $\bm\psi_0$ denotes the true value of $\bm\psi$, and $\bm f$ is the vector of estimating functions associated with $\bm\psi$. The explicit expression of $\bm f$ is provided in the Online Supplementary Materials. 

As shown in (\ref{sw}), estimation of the asymptotic variance $\V(\hat{\bm\psi})$ requires consistent estimators of the matrices
\[A(\bm\psi_0):=\E\bigg[-\frac{\partial \bm f}{\partial\bm\psi}(\bm O, \bm\psi_0) \bigg] \qquad \mathrm{and} \qquad B(\bm\psi_0):=\V\big[\bm f(\bm O, \bm \psi_0)\big].\]
Proposition \ref{prop:1} below provides an important insight into the feasibility of estimating these quantities under the current data structure.
\begin{proposition} \label{prop:1}
The matrix $A(\bm\psi_0)$ only involves terms of the form $\E\{I(S=k) \bm g(\bm O,\bm\psi_0)\}$, for $k=Z+1,\ldots,K$. \\
The matrix $B(\bm\psi_0)$ involves terms of the same form, as well as $\E(\bm L\bm L^\top|S=0)$, $\E(\bm L\bm L^\top|S=j)$ and $\E(\bm L Y|x,S=j)$, where $x=0,1$ and $j=1,\ldots,Z$.
\end{proposition}
A proof for Proposition \ref{prop:1} can be found in the Online Supplementary Materials. Importantly, this proposition implies that the sample analogue of $A(\bm\psi_0)$, obtained by replacing $\bm\psi_0$ with its estimates $\hat{\bm\psi}$, i.e., \[\frac{1}{n_0+n}\sum_{i=1}^{n_0+n}\frac{\partial f}{\partial\bm\psi}(\bm O_i, \hat{\bm\psi}),\] can be easily computed by just using data of trial $S=Z+1,\ldots, K$ with IPD. 

In contrast, estimation of $B(\bm\psi_0)$ requires terms such as $\E(\bm L \bm L^\top \mid S=0)$, $\E(\bm L \bm L^\top \mid S=j)$, and $\E(\bm L Y \mid X=x, S=j)$ for $j=1,\ldots,Z$. Because only AD are available for these populations, the sample analogues of these expectations cannot be directly computed. 
However, as suggested by equations (\ref{mod:mm2}) and (\ref{mod:mm}), all such nuisance parameters can be consistently estimated via inverse-probability weighting, without imposing additional assumptions on the data-generating process. Specifically, one can estimate the second-moment matrices of covariates $E(\bm L \bm L^\top|S=s);s=0,\ldots,Z$ by:
\begin{align} \label{est:ll}
    \hat \E(\bm L \bm L^\top\mid S=s) = \frac{1}{n_s}\sum_{k=Z+1}^{K} \bigg\{w_k\sum_{i:S_i=s} \bm L \bm L^\top m(\bm L, \hat{\bm\beta}_{sk}) \bigg\}
\end{align}
and the cross-moments with outcome $\E(\bm L Y|x,S=j);j=1,\ldots,Z$ by:
\begin{align} \label{est:ly}
\hat \E(\bm L Y\mid x,S=j) = \frac{1}{n_j}\sum_{k=Z+1}^{K} \bigg\{w_k\sum_{i:S_i = k}\bm L_i\cdot q(x, \bm L_i, \hat{\bm \phi}_j^{*(k)}, \hat{\bm \phi}_c^{(k)})\cdot m(\bm L_i,\hat{\bm\beta}_{jk})\bigg\} 
\end{align}
where $w_k$ are deterministic weights for the IPD trials, satisfying $\sum_{k>Z} w_k = 1$, and $m(\cdot)$ and $q(\cdot)$ are the weight and outcome functions defined previously. 

Equations (\ref{est:ll}--\ref{est:ly}) therefore provide a feasible strategy to recover all components needed for $\hat B$, enabling valid estimation of the asymptotic variance of $\hat{\bm\psi}$ by a sandwich estimator, using only IPD from trials $k>Z$ and available summary statistics from the AD trials and target population.

For example, one simple approach is to simulate pseudo-individual data for each AD trial $S=1,\ldots,Z$ and for the target population $S=0$, such that the summary statistics of these simulated data match the estimates \[\hat \E(\bm L \bm L^\top|S = s);s=0,\ldots,Z \qquad \text{ and } \qquad  \hat \E(\bm L Y|x,S=j); j=1,\ldots,Z\] obtained from (\ref{est:ll}) and (\ref{est:ly}), as well as other available summary statistics from the AD trials and target population (see Table \ref{tab:1}). Once constructed, standard software routines for sandwich variance estimation can be applied to these pseudo-data, toghether with the real IPD from trials $S=Z+1, \ldots, K$ to estimate $\V(\hat{\bm\psi})$. 

Importantly, this procedure remains valid even if the pseudo-IPD do not exactly replicate the true individual-level data, provided that the summary statistics above are preserved. Validity also relies on correct specification of the outcome model (\ref{mod1}) and the weight model (\ref{mod:we}), which are also required for consistency of the estimates $\hat\theta_s^0$. Beyond these model assumptions, no further assumptions are needed for asymptotically valid variance estimation.

\section{A simulation study}
\subsection{Simulation setup}
We conduct a simulation study to evaluate the finite-sample performance of the proposed approach, then compare its performance with other approaches that combine IPD and AD in the literature. Individual-level data from $K=5$ hypothetical trials $(S=1,\ldots,5)$ and a target population $(S=0)$ are generated as follows:
\begin{align*} \footnotesize
    &L_1 \sim Unif(0,1); \quad L_2 \sim Bern(0.5);
    \quad X \sim Bern(0.5) \\
    &\P(S=s|L_1,L_2) = \P(S=0|L_1,L_2)\cdot\exp(\bm\beta_s^\top \bm L)\quad\mathrm{for}\quad s=1,\ldots, 5\\
    &\P(S=0|L_1,L_2) = \bigg[1 + \sum_{s=1}^5\exp(\bm\beta_s^\top \bm L)\bigg]^{-1}\\
    &\P(Y=1|X,\bm L,S=s) = \mathrm{expit}(\phi_{0s} + \phi_{1s}X - 1.5L_1 + 1.5L_2 + \phi_{int}XL_2) \quad\mathrm{for}\quad s=1,\ldots, 5,
\end{align*}
where $\phi_{int} = 0.75$ and:
\begin{align*} 
\bm\beta_1=\bm\beta_2 =\bm \beta_3 &= \begin{pmatrix}
    0.15 & -0.10& -0.10
\end{pmatrix}^\top\\
\bm\beta_4 = \bm\beta_5 &= \begin{pmatrix}
    -0.15 & 0.10 & 0.10
\end{pmatrix}^\top\\
\begin{pmatrix}
    \phi_{01} & \ldots & \phi_{05}
\end{pmatrix}^\top &= \begin{pmatrix}
    0.25 & 0.50 & 0.25 & 0.25 & 0.50
\end{pmatrix}^\top\\
     \begin{pmatrix}
    \phi_{11} & \ldots & \phi_{15}
\end{pmatrix}^\top &= \begin{pmatrix}
    1.00 & 0.50 & 0.00 & 0.50 & 1.00
\end{pmatrix}^\top
\end{align*}
The summary statistics (D1) to (D4) but not the individual-level data are accessible in study $S=1,\ldots,3$ and in the target population $S=0$.
\subsection{Estimation strategies}
We first apply the proposed approach to meta-analyze results of the five trials after being standardized over the case-mix of the target population $S=0$. 
\begin{itemize}
    \item \textbf{Stage 1}: A correctly specified logistic regression model is fitted separately to the IPD from studies $S=4,5$ \answer{using standard maximum likelihood estimation}, yielding valid estimates of the coefficients for $L_1$, $L_2$ and the interaction term $XL_2$ that are assumed to be homogeneous across studies. The weight model (\ref{mod:we}) is then fitted using the proposed moment-based approach for each pair $(j,k)$, with $j=1,2,3$ and $k=4,5$, \answer{via the function \texttt{nleqslv()} from the \texttt{R} package \texttt{nleqslv}}. Estimates $(\hat\phi^{(k)}{0j}, \hat\phi^{(k)}{1j})$ of $(\phi_{0j}, \phi_{1j})$ are subsequently obtained by solving equation (\ref{eq5}) using the IPD from trial $k=4,5$, \answer{again via the same package}.
    \item \textbf{Stage 2}: The weighting approach based on (\ref{mod:tpe}) is applied to obtain valid estimates for the marginal treatment effects $\theta^0_s$ of each trial $s=1,\ldots,5.$ 
    \item \textbf{Stage 3}: The random-effect meta-analysis model (\ref{mod:re}) is fitted at stage 3 by using Markov Chain Monte Carlo (MCMC) simulations. \answer{  We adopt a Bayesian estimation framework due to its straightforward implementation using standard statistical software, which easily accommodates the potentially non-diagonal covariance structure of the standardized effects $\hat{\bm\theta}$. However, a frequentist approach, such as restricted maximum likelihood (REML), could be used equally well to fit model (\ref{mod:re}).} Weakly informative, independent prior distributions are assigned for the target parameters $(\Theta,\zeta)$, i.e. $\Theta \sim \mathcal{N}(0,10^6)$, and $\zeta \sim Unif(0,10)$. Posterior summaries are reported as the empirical medians of $\Theta$ and $\zeta^2$. 

    \answer{The MCMC simulations are implemented using the function \texttt{jags.model()} from the \texttt{R} package \texttt{rjags}, with 2 parallel chains and 10000 iterations for adaptation. To estimate the posterior medians of $\zeta^2$ and $\Theta$, 1000 samples per chain are drawn from the posterior distributions with a thinning interval of $5$, using function \texttt{coda.samples()} from \texttt{rjags}. The \texttt{R} codes implementing this analysis is available in the Online Supplementary Materials.} 
\end{itemize}
Three additional approaches are also applied to analyze the simulated data.
\begin{itemize}
    \item In the \textit{Ignoring-AD} approach, the AD studies $S=1,2,3$ are ignored, and only IPD studies $S=4,5$ are meta-analyzed by a two-stage strategy. In the first stage, a correctly specified logistic regression model is fitted separately to the IPD from each study. In the second stage, the treatment and interaction coefficient estimates from these two studies are averaged to obtain the summary treatment effects conditional on $\bm L$. \answer{This approach mirrors the standard two-stage IPD meta-analysis commonly used in practice, but it does not aim to transport trial results to a new target population.}
    \item In the \textit{classical multilevel-modeling} (Classical-ML) approach, we apply the strategy proposed by \citet{sutton2008meta} to combine the AD and IPD studies. Specifically, the following model is fitted to the simulated data:
\begin{align*}
    \E(Y|X, L_1, L_2, S=k) &= \mathrm{expit}\{\phi_{0k} + \phi_{1k}X + \phi_{2k}L_1 + \phi_{3k}L_2 + \phi_{int} XL_2\}\\
    \E(Y|X, S=j) &= \mathrm{expit}\{\phi_{0j} + \phi_{1j}X + \phi_{int}X\E(L_2|S=j)\}\\
    \phi_{1s} &\sim \mathcal{N}({\overline\phi}_1, \tau^2)
\end{align*}
where $j=1,2, 3$; $k = 4,5$ and $s = 1,\ldots,5$. A Bayesian MCMC approach is used to estimate this model. Following \citet{sutton2008meta}, we assign vague priors $\mathcal{N}(0,10^6)$ to the nuisance parameters $(\phi_{0k},\phi_{2k},\phi_{3k})$, and weakly informative, independent priors to the target parameters $(\overline{\phi}_1, \phi_{int}, \tau)$: specifically, $\overline{\phi}_1, \phi_{int} \sim \mathcal{N}(0,10^6)$ and $\tau \sim \mathrm{Uniform}(0,10)$. Posterior distributions of $\overline{\phi}_1$ and $\phi_{int}$ are summarized using the posterior median. \answer{Similar to the \textit{Ignoring-AD} approach, this method only estimates the conditional treatment effect given $\bm L$, without directly transporting trial results to the target population.}
\item In the \textit{multi-level network meta regression} (ML-NMR) approach, we follow the strategy proposed by \citet{phillippo2020multilevel} to combine AD and IPD studies. Pseudo-IPD are first generated for the AD trials $S=1,2,3$ using the available summary statistics, assuming a gamma distribution for $L_1$ and a Bernouilli distribution for $L_2$. An individual-level outcome model is then postulated and fitted across all studies, using the \texttt{R} package \texttt{multinma}:
\begin{align*}
    \E(Y|X, L_1, L_2, S=k) &= \mathrm{expit}\{\phi_{0k} + \phi_{1}X + \phi_{2}L_1 + \phi_{3}L_2 + \phi_{int} XL_2\}
\end{align*}
Model parameters are estimated via MCMC simulation, using weakly informative priors $\mathcal{N}(0,100)$ for all regression coefficients. Posterior inference for the treatment effect parameters $(\phi_1,\phi_{{int}})$ is based on their posterior medians. 

\answer{As noted in Remark~\ref{remark2}, transporting trial results to a new target population $S=0$ within the ML-NMR framework requires knowledge of the intercept $\phi_{00}=\E(Y \mid X=0,\bm L=\bm 0,S=0)$ in this population. In the present setting, however, only the marginal means and variances of $(L_1,L_2)$ are assumed to be available for $S=0$, rendering such transport infeasible. Consequently, ML-NMR yields only a summary estimate of the conditional treatment effect given $\bm L$, similarly to the \textit{Classical-ML} and \textit{Ignoring-AD} approaches.}
\end{itemize}
We consider two sample sizes, i.e. $n=2500$ and $n=5000$, and conduct $5000$ simulations for each sample size.

 \subsection{Analysis}
To evaluate our proposed approach, we first assess the absolute bias in estimating key parameters from stages 1 and 2. This is done by computing the mean difference between the estimated and true values of the parameters across simulation runs. At stage 1, key parameters include the trial-specific coefficients $(\phi_{0j},\phi_{1j})$, the average of the main effect of the treatment in the individual-level outcome model $\overline{\phi}_1=\sum_{j=1}^K\phi_{1j}/K$ and the interaction coefficient $\phi_{int}$. At stage 2, evaluated parameters include  the standardized treatment effects $\theta_s^0$, where $s=1,\ldots,5$. The true values of $\theta_0^s$ are obtained via a separate, large-scale simulation with $10^7$ runs, using the true model coefficients. For instance, the true value of $\theta_0^1$ is computed as $\log(\P_{1,j=1}^{0}/(1-\P_{1,j=1}^{0})) - \log(\P_{0,j=1}^{0}/(1-\P_{0,j=1}^{0}))$, where:
\[\P_{x,j=1}^{0} = \frac{\sum_{i=1}^{10^7}I(S_i=0)\mathrm{expit}(\phi_{00} + \phi_{10}x - 1.5L_1 + 1.5L_2 + \phi_{int}xL_2)}{\sum_{i=1}^{10^7}I(S_i=0)}\]
Across simulations, the variances of the key parameters at stages 1 and 2 are also estimated by using the procedure outlined in Table \ref{tab:1}. This enables one to establish a 95\% Wald confidence interval for each key parameter. The coverage of these intervals is subsequently assessed across simulations.

At stage 3, we evaluate the absolute bias in estimating 
$\Theta$ and $\zeta^2$ by calculating the mean difference across simulations between the posterior median of these parameters and their true values.  The true value of $\Theta$ is computed as $\sum_{s=1}^K\theta_s^{0}/K$ and of $\zeta^2$ is computed as $\sum_{s=1}^K(\theta_s^{0} - \Theta)^2/K$.

To evaluate the other three approaches (which focus on estimating the conditional treatment effect given $\bm L$), we compute the absolute bias as the average across simulations of the difference between the posterior medians of $(\overline\phi_1,\phi_{int})$ and their true values. \answer{As noted earlier, these approaches do not allow for transportation of trial results to a new target population. Consequently, direct comparison of these methods with stage 3 of the proposed approach is not feasible.}

\subsection{Simulation results}
Results of the simulation study can be found in table \ref{tab:2}. The proposed weighting-based approach correctly estimates the trial-specific coefficients $(\phi_{0j},\phi_{1j})$ in the AD studies $j=1,2,3$, the average main treatment effect $\overline{\phi}_1$ and the interaction effect $\phi_{int}$. The suggested procedure for variance estimation also performs well, which leads to correct coverage of the resulting 95\% confidence interval. 

As anticipated, both \textit{Ignoring-AD} and \textit{Classical-MM} approaches yield a biased estimate of 
$\overline{\phi}_1$, with comparable bias magnitude.  This bias arises from neglecting the AD trials in the \textit{Ignoring-AD} approach and from misspecification of the aggregate outcome model (i.e. ecological bias) in the \textit{Classical-MM} approach.

ML-NMR also yields biased estimates for both $\overline{\phi}_1$ and $\phi_{int}$. Compared to \textit{Classical-ML}, the bias magnitude when using ML-NMR is similar for $\overline{\phi}_1$, but 2 to 3 times larger for $\phi_{int}$. These biases arise from neglecting the heterogeneity in the main treatment effect $\phi_{1s}$ in the outcome model across studies, as well as from the use of an incorrect distribution to regenerate $L_1$ in the AD studies $S=1,2,3$.
\answer{\subsection{Finite-sample behaviors under model mispecification and poor overlap}}
\answer{To assess the sensitivity of the proposed approach to model mispecification and limited overlap between populations, we consider three additional simulation studies. In the first study, we induces poor overlap by modifying the numerical setup of the coefficients $\bm{\beta}$ in the weight model such that individuals with extreme covariate values are present only in a subset of trials. In the second study, we introduce misspecification of the weight model by including an additional term, $L_2^2$, into the true weight model, while omitting this term when estimating the weights at stage 1. Finally, in the third study, we use the same data generating mechanism as in the main simulation study but omit the interaction term when fitting the outcome model at stage 1. Under these assumption-violating settings, the proposed approach yields biased estimates of the outcome-model parameters $(\phi_{0j}, \phi_{1j})$ for the AD trials $j = 1, 2, 3$, which leads to biased marginal treatment effect estimates at subsequent stages of the analysis. Additional details on the numerical setups and the corresponding simulation results are provided in the Online Supplementary Materials.}

\section{Data analysis}

In this section, we apply the proposed method to compare the effectiveness of two psoriatic treatments--guselkumab and adalimumab--using real-world data from randomized controlled trials. Guselkumab is a newer interleukin-23 inhibitor (IL-23i), whereas adalimumab is a more established therapy that targets tumor necrosis factor-alpha (TNF-$\alpha$). Our analysis incorporates data from three trials: XPLORE (\citet{gordon2015phase}; trial 1), VOYAGE1 (\citet{blauvelt2017efficacy}; trial 2) and VOYAGE2 (\citet{reich2017efficacy}; trial 3) (table \ref{tab:3}). The trials differ in the dosing regimen of guselkumab. In XPLORE, patients received two 100 mg subcutaneous injections at baseline and week 8. In contrast, in VOYAGE1 and VOYAGE2, patients received 100 mg at baseline, week 4, and week 12.  This discrepancy in dosing schedules underscores potential heterogeneity in the treatment version across trials. 

For illustrative purposes, we consider the patient population observed in an independent trial, UltIMMa-1 \citep{gordon2018efficacy}, to be the target population. Data on three covariates, namely age $(L_1)$, PASI score at baseline $(L_2)$ and gender $(L_3)$ are collected across studies and in the target population. 

%To estimate the conditional treatment effect given covariates, a logistic outcome model is postulated in each trial. The linear predictor of this model includes treatment status, all covariates and an interaction between treatment status and sex. Table \ref{tab:4} shows the results when this logistic outcome model is separately fitted on the IPD of each study.

The outcome of interest is a binary indicator of whether a patient achieves at least a 90\% improvement in the Psoriasis Area and Severity Index (PASI) score at week 16 compared to baseline. This outcome is assumed to obey a logistic model, which includes treatment status, all baseline covariates, and an interaction between treatment status and gender.
The coefficients of all baseline covariates and of the treatment-gender interaction are homogeneous across studies, while the intercept and the main treatment effect are allowed to vary by trial to account for between-study heterogeneity. 

Suppose now that individual participant data (IPD) are available for trials 2 and 3, while only standard summary statistics (D1)-(D4) are available for trial 1 and the target population (as described above). To implement the proposed method under this data structure, stage 1 begins by fitting the specified logistic outcome model separately to the IPD from trials $S = 2$ and $S = 3$. Next, the weights $\P(S = 1 \mid \bm L) / \P(S = k \mid \bm L)$ for individuals in trial $k = 2, 3$ are estimated by fitting a logistic regression model that adjusts for all baseline covariates, using the proposed moment-based technique. Finally, equation (5) is solved twice, once for each source of IPD, to obtain two estimates of the treatment coefficient $\phi_{11}$ in trial 1, denoted by $\hat{\phi}_{11}^{(2)}$ and $\hat{\phi}_{11}^{(3)}$, using the appropriately weighted IPD from trials 2 and 3, respectively.

At stage 2, the weights $\P(S = 0 \mid \bm L) / \P(S = k \mid \bm L)$ for individuals in trials $k = 2, 3$ are estimated by fitting a logistic regression model that includes all baseline covariates as predictors. Once the weights are obtained, the sample analogue of equation (8) is solved for each combination of $(x, j, k)$, where $x \in \{0,1\}$, $j = 1$, and $k \in \{2,3\}$, yielding estimates of the standardized probabilities $\P^0_{xj}$. These standardized probabilities are then used to derive two separate estimates, $\hat{\theta}_1^{0,k=4}$ and $\hat{\theta}_1^{0,k=5}$, for the standardized treatment effect $\theta_1^0$. The final estimate of $\theta_1^0$ is obtained by averaging these two trial-specific estimates.

In parallel, for each of the trials   $k = 2, 3$, a standardized treatment effect estimate $\hat{\theta}_k^0$ is computed by solving the sample analogue of equation (8), using the outcome model fitted on the individual participant data (IPD) from the respective trial. 

At stage 3, the random effect model (\ref{mod:re}) is fitted by using MCMC simulations. The prior distribution of $\Theta$ is $\mathcal{N}(0,100)$, and of $\zeta^2$ is $\Gamma(1,2)$. The MCMC simulations are conducted with the \texttt{rjags} package in \texttt{R}, employing the same configuration as in the simulation study. The \texttt{R} code for implementation of this analysis can also be found on \href{https://github.com/votatthang/osma}{https://github.com/votatthang/osma}.

Resutls of this illustrative meta-analysis are presented in table \ref{tab:4}. The main treatment effect $\phi_{1s}$ is statistically significant in studies 2 and 3, but not in study 1. Correspondingly, the standardized treatment effect $\theta_2^0$ and $\theta_3^0$ are also statistically significant, indicating that the versions of guselkumab evaluated in these trials outperform standard adalimumab in the target population. In contrast, no statistical difference is observed between the version of guselkumab evaluated in trial 1 and adalimumab. Furthermore, a high between-study variance is observed in stage 3 of the analysis, suggesting substantial heterogeneity across trials even after adjusting for case-mix differences. Such heterogeneity is potentially overlooked when the difference in dosing schedule between trials is ignored.

\section{Discussion}
In this paper, we introduce a new method to incorporate aggregate data into a causally interpretable meta-analysis of individual participant data. The proposed method employs inverse weighting to estimate trial-specific coefficients in the (individual-level) outcome model of studies for which IPD are unavailable. In contrast to hierarchical modeling approaches which typically require specifying a model that linkes aggregate outcomes and aggregate covariates--a task that is particularly challenging in non-linear settings--our method provides a simpler and more flexible alternative. Moreover, the proposed framework can be readily extended to incorporate non-randomized studies. 

\answer{As discussed earlier, limited covariate overlap between the trial populations and the target population may lead to extreme weights, which can destabilize the proposed method. In such settings, ad hoc strategies such as weight truncation or trimming may be used to mitigate the influence of extreme weights \citep{remiro2022two}. However, these approaches are often sensitive to the choice of cutoff values and may discard a substantial portion of the data, potentially compromising efficiency and interpretability. Future research should therefore investigate whether more principled methods developed in the causal inference literature to handle extreme weights such as overlap weighting or covariate balancing approaches can be effectively integrated into the proposed framework \citep{matsouaka2020framework, li2019addressing}. In addition, to reduce the risk of model misspecification in both the outcome and weight models, semi-parametric and nonparametric (data-adaptive) modeling strategies could be incorporated within the proposed framework.}

It is also worth investigating the performance of the proposed approach under settings with varying numbers and relative sizes of studies for which IPD are available, as well as extending the method to network meta-analysis, where the practical challenges posed by inaccessible IPD may be even more pronounced. Finally, in many applications, different trials collect data on different sets of patient characteristics \citep{vuong2024development}. Extending the proposed method to accommodate partially overlapping covariate sets would substantially enhance its applicability in real-world evidence synthesis.

\section{Competing interests}
No competing interest is declared.

\section{Author contributions statement}
T.T.V developed the proposal. T.T.K.L, M-F.B and T.T.V conducted the simulation study, T.T.K.L and S.A conducted the illustrative data analysis. T.T.K.L., M-F.B and T.T.V wrote and reviewed the manuscript.

\section{Acknowledgments}
T.T.V is supported by the French National Research Agency (Agence Nationale de la Recherche), through a funding for Chaires de Professeur Junior (23R09551S-MEDIATION).

\section*{Data availability}
The clinical trial data that support the findings in this paper are from the Vivli platform. More information on how to get access to these
data is available on the platform's website (\href{https://vivli.org/}{https://vivli.org/}).

\bibliographystyle{plainnat}
\bibliography{reference}
\newpage
\section*{Tables and figures}
\begin{table}[tbh!]
%\captionsetup{margin=1.2cm, font={stretch=1.5}}
\centering
\caption{Estimating the asymptotic variance of $\hat{\bm\psi}$, the proposed estimator for the target parameter $\bm\psi$. (*): The simulated outcome $Y^*$ in this procedure is no longer binary, but it does not affect the validity of the output.}
\scalebox{1}{\begin{tabular}{ p{1cm} p{15cm}}
\hline
Step & Procedure\\[2pt] \hline
1. & For $j=1,\ldots, Z$, estimate $\E(\bm T\bm T^\top|x,j)$ by (\ref{est:ll}) and (\ref{est:ly}), where $\bm T = \begin{pmatrix}
    \bm L^\top & Y
\end{pmatrix}^\top$. Denote $\hat \E(\bm T\bm T^\top|x,j)$ the obtained estimate.\\[2pt]
2. & Calculate \[\hat\V(\bm T|x,j) = \hat \E(\bm T\bm T^\top|x,j) - \hat \E(\bm T|x,j)\hat \E(\bm T|x,j)^\top\]
where $\hat \E(\bm T|x,j)$ is the sample average of $\bm T$ in treatment group $X=x$ in study $S=j$, which is available in the summary statistics of this trial. \\[2pt]
3. & Simulate pseudo-individual level data $\bm O_{ji}^* = \begin{pmatrix}
    X_{ji}^* & \bm T_{ji}^{*,\top}
\end{pmatrix}^\top$ where $i=1,\ldots, n_j$ for each trial $j=1,\ldots,Z$, as follows:
\begin{enumerate}
    \item[3a.] Generate the (pseudo-) treatment status as $X_{ij}^* = I(i\le n_{xj}).$
    \item[3b.] Simulate $\bm T^*_{ij}$ such that the sample mean and the sample covariance matrix of the subgroup with $X_{ij}^*=x$ match $\hat \E(\bm T|x,j)$ and $\hat\V(\bm T|x,j)$. This can be done in \texttt{R}, for instance, by using the function \texttt{mvrnorm()} (package \texttt{MASS}) with \texttt{empirical = TRUE}.$^*$      
\end{enumerate}
\\[-5pt]
4 & Estimate $\V(\bm \psi)$ by a sandwich estimator, using the simulated data from trial $S=1, \ldots, Z$ and the real individual-level data from trial $S=Z+1,\ldots, K$.\\[2pt]
5 & Apply the Delta-method to obtain the asymptotic variance of $\hat\theta_s^{0}$, for $s=1,\ldots, K$. \\
\hline
\end{tabular}}
\label{tab:1}
\end{table}

\begin{center}
    \begin{table}[H]
    %\captionsetup{margin=1.2cm, font={stretch=1.5}}
    \centering
    \caption{Simulation results. PAPS: the proposed approach of Population Adjustment based on Propensity Score. Classical-MM: classical multilevel modeling approach. Ignoring AD: standard IPD meta-analysis which ignores studies with aggregated data. ML-NMR: Multilevel Network Meta-Regression.}
    \label{tab:2}
    \scalebox{0.7}{\begin{tabular}{cclrrrr}
    \hline
    Approach & Sample size & Parameter & Bias & $\V(\hat\phi)$ & $\hat\V(\hat\phi)$ & Coverage (\%) \\
    \hline
    \multirow{20}{*}{PAPS} & \multirow{20}{*}{$5000$} & \textbf{Stage 1}\\
    & & $\phi_{01}$  & $3.5\times 10^{-3}$ & $2.76\times 10^{-2}$ & $2.77\times 10^{-2}$ & 95.4 \\
    & & $\phi_{02}$     & $4.6\times 10^{-3}$ & $2.81\times 10^{-2}$ & $2.77\times 10^{-2}$ & 95.4\\
    & & $\phi_{03}$ &  $4.6\times 10^{-3}$ & $2.79\times 10^{-2}$ & $2.77\times 10^{-2}$ & 95.0\\
    & & $\phi_{11}$     & $9.3\times 10^{-3}$ & $4.23 \times 10^{-2}$ & $4.16\times 10^{-2}$ & 94.9\\
    & & $\phi_{12}$     & $7.0\times 10^{-3}$ & $4.06\times 10^{-2}$ & $4.07\times 10^{-2}$ & 95.3\\
    & & $\phi_{13}$ & $-4.7\times 10^{-3}$ &$4.39\times 10^{-2}$ & $4.42\times 10^{-2}$ & 95.2\\
    & & $\overline{\phi}_1$ & $ 5.1\times 10^{-3}$ & $1.57\times 10^{-2}$ & $1.54\times 10^{-2}$ & 94.9  \\
    & & $\phi_{int}$ &  $0.6\times 10^{-2}$ & $7.94\times 10^{-2}$ & $7.63\times 10^{-2}$ & 95.5\\
    & & \textbf{Stage 2}\\
    & & $\theta_{1}^{0,4}$ & $3.7\times 10^{-3}$ & $2.56\times 10^{-2}$  & $2.56\times 10^{-2}$ & $95.1$\\
    & & $\theta_{1}^{0,5}$ & $3.4\times 10^{-3}$ & $2.56\times 10^{-2}$ & $2.54\times 10^{-2}$ & 95.2 \\
    & & $\theta_{2}^{0,4}$ & $3.9\times 10^{-3}$ & $2.39\times 10^{-2}$ & $2.46\times 10^{-2}$ & 95.6 \\
    & & $\theta_{2}^{0,5}$ & $4.1\times 10^{-3}$ & $2.41\times 10^{-2}$ & $2.46\times 10^{-2}$ & 95.2 \\
    & & $\theta_{3}^{0,4}$ & $-1.5\times 10^{-3}$ & $2.26\times 10^{-2}$ & $2.23\times 10^{-2}$ & 95.4 \\
    & & $\theta_{3}^{0,5}$ & $-1.0\times 10^{-3}$ & $2.27\times 10^{-2}$ & $2.23\times 10^{-2}$ & 95.5 \\
    & & $\theta_{4}^{0,4}$ & $1.0\times 10^{-3}$ & $1.94\times 10^{-2}$ & $1.87\times 10^{-2}$ & 94.6  \\
    & & $\theta_{5}^{0,5}$ & $4.3\times 10^{-3}$ & $2.42\times 10^{-2}$ & $2.39\times 10^{-2}$  & 95.2\\
    & & \textbf{Stage 3}\\
    & & $\Theta$& $-2.9\times 10^{-4}$ & - & - & -\\
    & & $\tau^2$& $7.2\times 10^{-2}$ & - & - & -\\
    \hline
    \multirow{2}{*}{Classical-ML} &  \multirow{2}{*}{$5000$} & $\overline{\phi}_1$  & $-1.3\times 10^{-1}$ & - & - & -\\
    & & $\phi_{int}$ & $-1.4\times 10^{-2}$ & - & - & -\\
    \hline
    \multirow{2}{*}{ML-NMR} & \multirow{2}{*}{$5000$} & $\overline{\phi}_1$  & $1.2\times 10^{-1}$ & - & - & -\\
    & & $\phi_{int}$ &  $-7.3\times 10^{-2}$ & - & - & -  \\
    \hline
    \multirow{2}{*}{Ignoring AD} & \multirow{2}{*}{$5000$} &   $\phi_{1}$ & $1.7\times 10^{-1}$ & $2.30\times 10^{-2}$ & $2.42\times 10^{-2}$ & 81.4   \\
    & & $\phi_{int}$ & $0.6\times 10^{-2}$ & $7.94\times 10^{-2}$ & $7.63\times 10^{-2}$ & 95.5\\
    \hline\hline
    \multirow{20}{*}{PAPS} & \multirow{20}{*}{$2500$} & \textbf{Stage 1}\\
    & & $\phi_{01}$  & $0.9\times 10^{-2}$ & $5.67\times 10^{-2}$ & $5.64\times 10^{-2}$ & 95.3\\
    & & $\phi_{02}$  & $1.5\times 10^{-2}$ & $5.92\times 10^{-2}$ & $5.64\times 10^{-2}$ & 94.7\\
    & & $\phi_{03}$ &  $1.4\times 10^{-2}$ & $5.85\times 10^{-2}$ & $5.63\times 10^{-2}$ & 94.8\\
    & & $\phi_{11}$  & $1.9\times 10^{-2}$ & $8.67\times 10^{-2}$ & $8.50\times 10^{-2}$ & 94.8\\
    & & $\phi_{12}$ & $0.1\times 10^{-2}$ & $8.53\times 10^{-2}$ & $8.27\times 10^{-2}$ & 94.9\\
    & & $\phi_{13}$ & $-1.0\times 10^{-2}$ & $9.38\times 10^{-2}$ & $9.00\times 10^{-2}$ & 94.9\\
    & & $\overline\phi_{1}$ & $0.7\times 10^{-2}$ & $3.28\times 10^{-2}$ & $3.15\times 10^{-2}$ & 94.6\\
    & & $\phi_{int}$ & $3.0\times 10^{-2}$ & $1.60\times 10^{-1}$ & $1.65\times 10^{-1}$ & 96.0\\
    & & \textbf{Stage 2}\\
    & & $\theta_{1}^{0,4}$ & $1.2\times 10^{-2}$ & $5.27\times 10^{-2}$ & $5.21\times 10^{-2}$ & 94.8 \\
    & & $\theta_{1}^{0,5}$ & $0.8\times 10^{-2}$ & $5.22\times 10^{-2}$ & $5.23\times 10^{-2}$ & 95.1  \\
    & & $\theta_{2}^{0,4}$ & $0.2\times 10^{-2}$ & $5.05\times 10^{-2}$ & $5.00\times 10^{-2}$  & 95.2\\
    & & $\theta_{2}^{0,5}$ & $-0.2\times 10^{-2}$ & $4.98\times 10^{-2}$ & $5.00\times 10^{-2}$  & 95.4 \\
    & & $\theta_{3}^{0,4}$ & $0.2\times 10^{-2}$ & $4.60\times 10^{-2}$ & $4.52\times 10^{-2}$ & 94.8 \\
    & & $\theta_{3}^{0,5}$ & $-0.2\times 10^{-2}$ & $4.66\times 10^{-2}$ & $4.54\times 10^{-2}$ & 94.5  \\
    & & $\theta_{4}^{0,4}$ & $0.6\times 10^{-2}$ & $3.82\times 10^{-2}$ & $3.76\times 10^{-2}$ & 94.8 \\
    & & $\theta_{5}^{0,5}$ & $0.8\times 10^{-2}$ & $4.92\times 10^{-2}$ & $4.81\times 10^{-2}$ & 94.8 \\
    & & \textbf{Stage 3}\\
    & & $\Theta$ & $-4.6\times 10^{-4}$ & - & - & -\\
    & & $\tau^2$ & $6.5\times 10^{-2}$ & - & - & -\\
    \hline
    \multirow{2}{*}{Classical-ML} &  \multirow{2}{*}{$2500$} & $\phi_{1}$  &  $-1.4\times 10^{-1}$ & - & - & -\\
    & & $\phi_{int}$ & $-2.3\times 10^{-2}$ & - & - & - \\
    \hline
    \multirow{2}{*}{ML-NMR} &  \multirow{2}{*}{$2500$} & $\phi_{1}$  & $1.3\times 10^{-1}$ & - & - & -\\
    & & $\phi_{int}$ &  $-4.7\times 10^{-2}$ & - & - & -\\
    \hline
    \multirow{2}{*}{Ignoring AD} & \multirow{2}{*}{$2500$} &   $\phi_{1}$  & $1.6\times 10^{-1}$ & $5.57\times 10^{-2}$ & $4.94\times 10^{-2}$ & 85.9\\
    & & $\phi_{int}$ & $3.0\times 10^{-2}$ & $1.60\times 10^{-1}$ & $1.65\times 10^{-1}$ & 96.0\\
    \hline
    \end{tabular}}
    \label{tab:2}
    \end{table}
\end{center}
\begin{center}
    \begin{table}[H]
    \centering
    \caption{Data analysis: baseline characteristics and outcome across studies and in the target population. PASI: Psoriasis Area and Severity Index.}
    \label{tab:3}
    %\resizebox{\textwidth}{!}{
    \scalebox{0.7}{\begin{tabular}{lrrrrrrr}
    \hline
    Characteristics& Target population & \multicolumn{2}{c}{Study 1} & \multicolumn{2}{c}{Study 2} & \multicolumn{2}{c}{Study 3}\\
    \hline
    Treatment & - & Guselkumab & Adalimumab & Guselkumab & Adalimumab & Guselkumab & Adalimumab \\
    \quad No. (\%) & 508 (100.0) & 202 (83.8)& 39 (16.2)& 323 (50.0) & 323 (50.0) & 481 (67.1) & 236 (32.9) \\
    \hline
    Covariates & & & & & & & \\
    \quad Men - no. (\%) & 363 (71.5) & 143 (70.8) & 26 (66.7)& 238 (73.7)& 241 (74.6) & 337 (70.1)& 162 (68.6) \\
    \quad Age - Mean (SD) & 48.1 (13.4) & 39.8 (12.6) & 40.3 (14.8) & 41.9 (12.8) & 41.0 (12.8) & 43.2 (12.2) & 42.8 (12.0) \\
    \quad PASI score week 0 - Mean (SD)& 20.4 (7.4) & 20.9 (8.1) & 20.2 (7.5) & 22.2 (9.5) & 22.3 (8.8) & 21.9 (8.6) & 21.6 (8.9)\\
    \hline
    Outcome & & & & & & &\\
    \quad PASI score week 16 - Mean (SD) & - & 3.9 (5.7) & 3.6 (5.1) & 1.7 (3.3) & 4.2 (6.5) & 2.0 (3.4) & 4.2 (6.3) \\
    \quad PASI 90 - no. (\%) & - & 99 (49.0) & 19 (48.7) & 242 (74.9)& 166 (51.4)& 345 (71.7)& 116 (49.2)\\
    \hline
    \end{tabular}}
    \label{tab:3}
    \end{table}
\end{center}

\begin{center}
    \begin{table}[H]
        \centering
        \caption{Data analysis: population-adjusted meta-analysis results. $\phi_{0j}$: intercept (specific to study $j$) in the logistic outcome model; $\phi_{0j}$: treatment coefficient (specific to study $j$) in the logistic outcome model; $\phi_{int}$: treatment-gender interaction coefficient (homogeneous across studies); $\theta_s^0$: effect of treatment version $s$ versus control version $s$ in the target population (log odds ratio scale); $\Theta$: summary treatment effect in the target population (log odds ratio scale); $\tau^2$: treatment version heterogeneity.}
        \label{tab:4}
        \begin{tabular}{l r r}
        \hline
        \textbf{Parameter} & \textbf{Estimate} & \textbf{95\% Confidence Interval} \\
        \hline
        \textbf{Stage 1}\\
        $\phi_{01}$     & $-0.033$ & $-0.866; 0.800$\\
        $\phi_{11}$     & $0.378$ & $-0.415; 1.171$\\
        $\phi_{02}$     & $0.509$ & $-0.329; 1.347$\\
        $\phi_{12}$     & $1.267$ & $0.597; 1.935$\\
        $\phi_{03}$     & $-0.421$ & $-1.232; 0.389$\\
        $\phi_{13}$     & $1.514$ & $ 0.906; 2.121$\\
        $\phi_{int}$    & $-0.473$ & $-1.002;0.056$ \\
        \hline     
        \textbf{Stage 2}\\
        $\theta_{1}^0$ & $0.050$ & $ -0.623; 0.723$\\
        $\theta_{2}^0$ & $1.101$ & $ 0.771; 1.430$\\
        $\theta_{3}^0$ & $0.999$ & $ 0.674; 1.323$\\
        \hline
        \textbf{Stage 3}\\
        $\Theta$ & $0.755$ & $ -0.333; 1.735$\\
        $\tau^2$ & $2.169$ & $ 0.307; 8.184$\\
        \hline
        \end{tabular}
    \end{table}
\end{center}

\end{document}